# Phase Diagram and Weak-link Behavior in Nd-doped $CaFe_2As_2$

Bo Gao, Xiaojiang Li, Qiucheng Ji, Gang Mu*, Wei Li, Tao Hu, Ang Li, and Xiaoming Xie

State key Laboratory of Functional Materials for Informatics, Shanghai Institute of Microsystem and Information Technology, Chinese Academy of Sciences, Shanghai 200050, China

**Abstract**

The transport properties, phase diagram, and dopant distribution are investigated in systematically Nd doped $CaFe_2As_2$ single crystals. Coexistence of two superconducting (SC) phases with different critical transition temperature ($T_c$) is observed. The low-$T_c$ phase emerges as $x \geq 0.031$, and the $T_c$ value increases to its maximum value of about 20 K at $x = 0.083$, the maximum doping level in our study. As $x \geq 0.060$, the high-$T_c$ phase with a $T_c$ value of about 40 K is observed. The structure transition (STr) from tetragonal to orthorhombic phase vanishes suddenly around $x = 0.060$, where a new STr from tetragonal to collapsed tetragonal phase begins to turn up. Compared to the low-$T_c$ phase, the end point of SC transition of the high-$T_c$ phase is more sensitive to the magnetic field, showing a characteristic of Josephson weak-link behavior. Possible scenarios about this system are discussed based on our observations. We also find that the non-uniform SC properties cannot be attributed to the heterogeneous Nd distribution on the micro scale, as revealed by the detailed energy dispersive X-ray spectroscopy (EDS) measurements.

Corresponding author:
E-mail address: mugang@mail.sim.ac.cn (Gang Mu)

## 1. Introduction

Fe-based superconductors have been studied extensively since the report of LaFeAsO$_{1-x}$F$_x$ with $T_c$ of 26 K [1, 2]. Among the different systems, the AFe$_2$As$_2$ compounds (A=Ba, Sr, Ca, Eu, so called "122" system) with the ThCr$_2$Si$_2$-type structure [3] are widely studied because single crystals with high quality are easily accessible [4]. The parent compounds of the 122 system undergo a phase transition from a high temperature tetragonal, paramagnetic phase (T phase) to a low temperature orthorhombic, antiferromagnetic phase (O phase). The antiferromagnetic order can be systematically suppressed and superconductivity can develop by the means of chemical substitution or applying pressure. The highest $T_c$ value in the 122 system is still lower than 55 K in the $R$FeAsO (1111) system [5]. Superconductivity with maximum $T_c$ of 38 K has been achieved in the Ba$_{1-x}$K$_x$Fe$_2$As$_2$ by hole-doping [6]. Meanwhile, electron-doping usually induces superconductivity at a lower temperature (around 22 K) by substituting Fe ions with other transition metals [7-9]. This is typically attributed to the imperfection of the FeAs conducting layer induced by doping.

In order to further enhance the $T_c$, much attention has been paid to electron doping approached by substitution of trivalent rare-earth elements ions (Re$^{3+}$) on divalent A$^{2+}$ ions in 122 system without affecting the FeAs layers [10-16]. However, superconductivity in single-crystalline samples is only attained in systems based on CaFe$_2$As$_2$. Besides the T-O transition at ambient pressure for CaFe$_2$As$_2$, the tetragonal phase transforms to a new collapsed tetragonal structure (cT, both the a-axis and c-axis lattice shrink) when a hydrostatic pressure (> 0.35 GPa) is applied [17, 18]. Recently, it is found that this cT phase can be stabilized at ambient pressures by doping Pr or Nd into CaFe$_2$As$_2$. In contrast, the substitution of up to 28% La or 17% Ce does not drive this T-cT transition [12]. More surprisingly, two superconducting phases with $T_c$ of about 20 K and 40-49 K respectively were discovered in the rare-earth doped Ca$_{1-x}$Re$_x$Fe$_2$As$_2$ (Re = La, Ce, Pr, Nd) compounds, regardless of this T-cT structural evolution [12-16]. Although the high-$T_c$ phase exceeds the highest $T_c$ ~ 38 K in the hole-doped Ba$_{1-x}$K$_x$Fe$_2$As$_2$, the superconducting volume fraction is very low suggesting the absence of bulk superconductivity. The origin of the non-bulk and two-phase superconductivity has be attributed to the minor foreign phase, interface or filamentary superconductivity, Josephson junction coupling between grains *et al*, which is still an open issue and needs more in-depth investigations [14, 19, 20].

To the best of our knowledge, a systematic investigation on the Nd-doped $CaFe_2As_2$ system is still lacking. Moreover, the temperature versus doping phase diagram of this system is still not clear. In the present work, we report a systematic investigation of the characterization and phase diagram of the electron-doped $Ca_{1-x}Nd_xFe_2As_2$ single crystals. The behaviors of field induced resistance broadening for superconducting transition are also observed, indicating a weak-link feature in the present system.

## 2. Experimental Details

Single crystals of systematically Nd-doped $CaFe_2As_2$ were grown using a self-flux method. The FeAs precursor was synthesized by the reaction of Fe powder and As chips at 700 ℃ for 20 h in a vacuum quartz tube. Appropriate amounts of the starting materials of FeAs, Ca and Nd with the ratio of 4: (1-*x*): *x* were placed in an alumina crucible, and sealed in an arc-welded iron tube. The sample was heated to 1200 ℃ slowly and held for 5 hours, and then cooled to 1030 ℃ with a rate of 3-6 ℃/hour to grow the single crystals. The obtained single crystals show a shiny surface and are easily cleaved into plates.

The phase identification and crystal structure were characterized by X-ray diffraction (XRD) with Cu K radiation. The actual Nd concentrations were checked and determined by the energy dispersive X-ray spectroscopy (EDS) measurements. The resistance measurements with magnetic fields up to 9 T were carried out by using a standard four-contact method with a quantum design physical property measurement system (PPMS).

## 3. Results and discussion

Figure 1a shows the XRD *θ-2θ* patterns for four typical $Ca_{1-x}Nd_xFe_2As_2$ single crystals with different doping levels. The sharp (00*l*) diffraction peaks suggest that the crystallographic *c*-axis is perpendicular to the plane of the single crystals with an excellent crystalline quality. The calculated *c*-axis lattice parameters as a function of Nd content are plotted in Figure 1b. The data of parent phase are taken from the report by S. R. Saha *et al.* [12]. It can be found that *c* axis shrinks monotonously with increasing *x*, which implies a successful chemical substitution and is also consistent with previous reports [12].

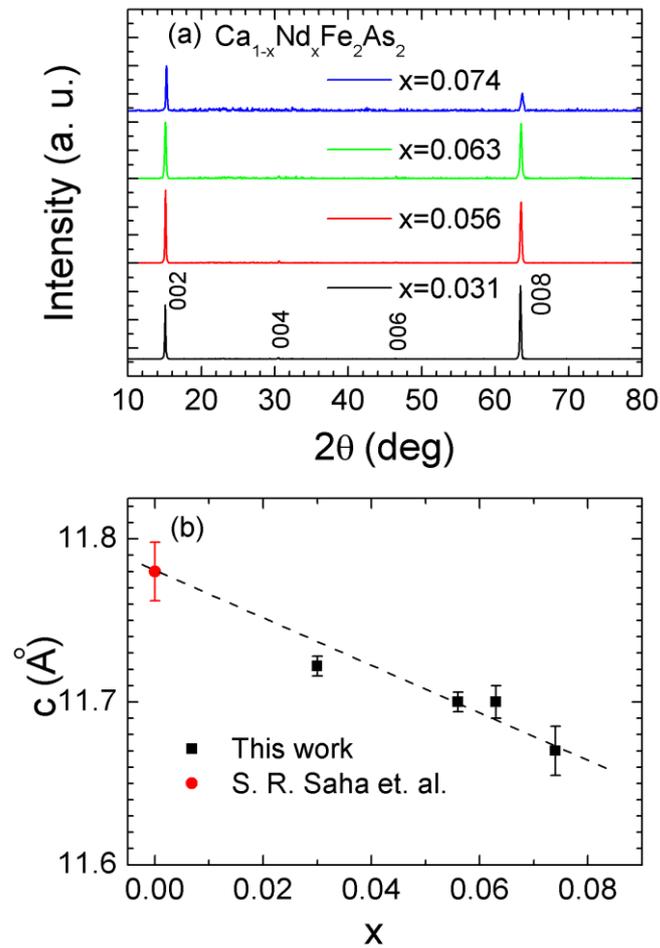

Figure. 1 (a) X-ray diffraction patterns of $Ca_{1-x}Nd_xFe_2As_2$ for different doping levels of x. (b) Doping dependence of *c*-axis lattice parameters. The dashed line is the guide for eyes. The data of parent phase are taken from the report by another group [12].

Figure 2 presents the temperature dependence of resistivity under zero fields for $Ca_{1-x}Nd_xFe_2As_2$ single crystals, normalized to the data at 300 K. The data of parent phase are taken from the report by another group [12]. Several features are observed at different temperatures. In the inset of figure 2, we denote them by arrows for the sample with $x = 0.060$ as an example, where $T_O$, $T_{cT}$, $T_{cH}$, and $T_{cL}$ represent the transition temperature to the orthorhombic phase, to the collapsed tetragonal phase, the onset transition temperature of high-$T_c$ phase, and that of the low-$T_c$ phase, respectively.

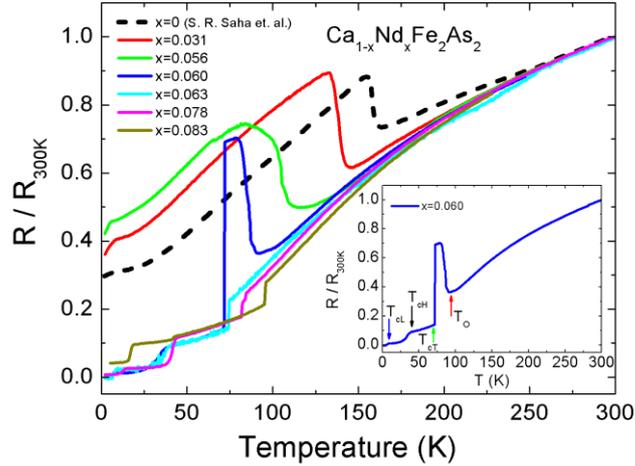

Figure. 2. Temperature dependence of the in-plane electrical resistivity for $Ca_{1-x}Nd_xFe_2As_2$ single crystals, normalized to the data at 300K. The data of parent phase are taken from the report by another group [12].

From our data we can see that with Nd doping the resistivity anomaly due to the tetragonal to orthorhombic STr shifts gradually to lower temperature and disappears around $x = 0.060$. Another conspicuous feature is a sharp and dramatic drop in resistivity when the doping level $x \geqslant 0.060$. This feature is associated with a STr from T phase to cT phase [12, 17]. We note that there exists a hysteresis for the T-cT STr with increasing and decreasing the temperature. Here we only show the data collected with increasing the temperature. With increasing $x$, this resistivity transition shifts to higher temperatures, which is similar to that observed in $Ca_{1-x}Pr_xFe_2As_2$ based on neutron-diffraction measurements [12]. For the sample with $x = 0.060$, the coexistence of two structure transitions may be due to the local inhomogeneity. Along with the suppression of T-O phase transition, resistivity decreases quickly below 10 K as $x \geq 0.031$, suggesting the appearance of a superconducting transition. When $x \geq 0.060$, two superconducting transition steps appear in low temperature region, which seems to be a common feature in $Ca_{1-x}Re_xFe_2As_2$. Both superconducting transitions are broad and no zero resistance was observed in some of the samples down to 2 K.

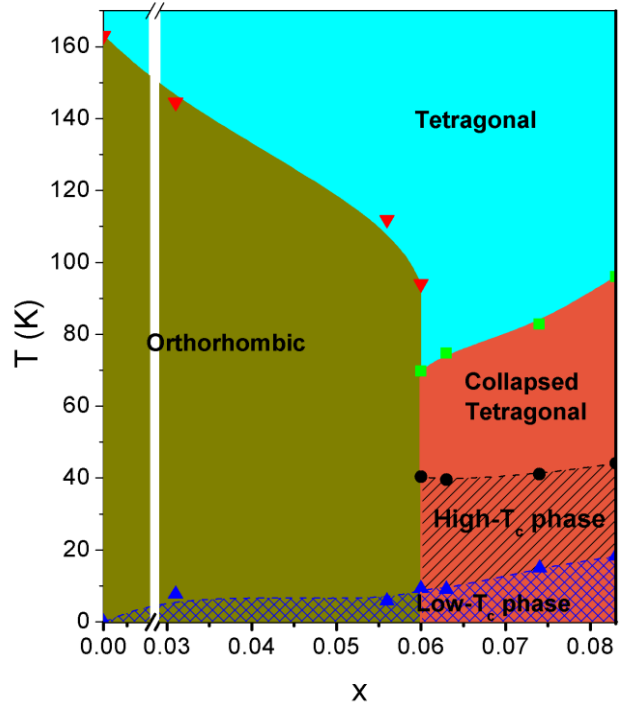

Figure. 3. Doping-temperature ($x$-$T$) phase diagram of $Ca_{1-x}Nd_xFe_2As_2$. The data of parent phase are taken from the report by another group [12]. The regions with different colors represent the different structural phases. The two SC phases with different $T_c$ are revealed by black and blue patterns, respectively.

Based on the resistivity behavior described above, we can establish a doping-temperature ($x$ - $T$) phase diagram for $Ca_{1-x}Nd_xFe_2As_2$, which is shown in figure 3. In the lower-doped side ($x \leqslant 0.060$), superconductivity of low-$T_c$ phase coexists with T-O transition. Similar behaviors have also been observed in the hole-doped $Ba_{1-x}K_xFe_2As_2$ [21], electron-doped $BaFe_{2-x}Co_xAs_2$ [22] and other rare-earth doped $CaFe_2As_2$ systems [13]. When the doping level increases to 0.060, the T-O STr vanishes suddenly and a new T-cT STr begins to turn up. At the same time, the high-$T_c$ phase emerges with an almost invariably $T_c$ value of about 40 K. In contrast, the $T_c$ value of the low-$T_c$ phase is monotonically increased from 10 K to about 20 K with $x$ increasing. Unlike other rare-earth elements doped CaFe$_2$As$_2$, both superconducting phases are detected clearly from the resistivity data in the high-doped range in our system, meaning that at least the high-$T_c$ phase doesn't form the continuous percolative path for the current. This may be due to the lower solubility limit of the Ca$_{1-x}$Nd$_x$Fe$_2$As$_2$ compounds. Nevertheless, the distinct coexistence of two SC transitions in resistivity facilitates our investigation on the intrinsic natures of the two SC phases (see the next paragraph). Compared our results with other reports on La and Pr-doped CaFe$_2$As$_2$ systems, it is concluded that the high-$T_c$ phase appears at the doping level when the structural (T-O)/antiferromagnetic phase transition is totally suppressed. This is further demonstrated recently by the high pressure work on Ca$_{1-x}$La$_x$Fe$_2$As$_2$ samples [23]. The high-$T_c$ phase only coexists with T-O transition in one sample with $x$ = 0.060 possibly due to the local inhomogeneity.

To check the influence of magnetic fields on the two superconducting phases, we measured the temperature dependence of the resistivity under different magnetic fields up to 9 T. The magnetic fields were applied along the $c$-axis of the single crystals. Here we show the data for one sample with $x$ = 0.060 (denoted as 0.060-2) in figure 4. The transition temperature of superconductivity is suppressed gradually and the transition is broadened with increasing the magnetic fields. However, obvious differences for the influence of magnetic field on the two SC phases are observed. An unconspicuous field induced resistance broadening behavior is observed in the low-$T_c$ phase. For the high-$T_c$ phase, in contrast, the end point of the SC transition is very sensitive to the magnetic field, which shifts obviously to lower temperatures even under a magnetic field of 0.05 T. We argue that this is a typical characteristic of Josephson weak links, which has been observed at the high-angle grain boundaries in high-$T_c$ cuprate superconductors [24, 25].

We attempted to further explore the possible origins of the non-uniform SC properties in the present system. The distribution of the Nd-dopant on micro-scale is investigated by EDS measurements. The mapping image of Nd concentrations and the chart of its distributions throughout an area of 36×26 μm for the sample with $x$= 0.074 are shown in figure 5(a) and (b). The spatial resolution is 1.9 μm. It shows a Nd distribution ranging from 0.061 to 0.086 and an average concentration of 0.074, which is similar to that reported on Pr-doped CaFe2As2 system [20]. The full width at half maximum (FWHM) of the profile for the histogram in figure 5(b) is about 0.009. Our data indicate that the two-SC-phase feature observed in the present system cannot be attributed to the Nd distribution on the micro scale. Of course we cannot rule out possible heterogeneous features responsible for the non-uniform SC behaviors on a smaller scale (e.g. nano scale). It was indicated that the high-$T_c$ phase is not an interfacial superconductivity [23] or a filamentary-type superconductivity caused by local pinning strength and local structural defects [12, 13]. Very recently, K. Gofryk et al. [20] reported that the inhomogeneous and strongly localized high-$T_c$ phase is a kind of granular filamentary superconductivity emerging from clover-like regions associated with Pr dopants composed of 3 or 4 atoms in Pr-doped $CaFe_2As_2$. These regions with a SC gap of $\Delta \sim 30$ meV are separated and surrounded by other low-$T_c$ phases with $\Delta \sim 15$ meV and non-SC regions. So the weak-link behavior of high-$T_c$ phase observed in our data is likely to originate from the boundaries between these high-$T_c$ regions.

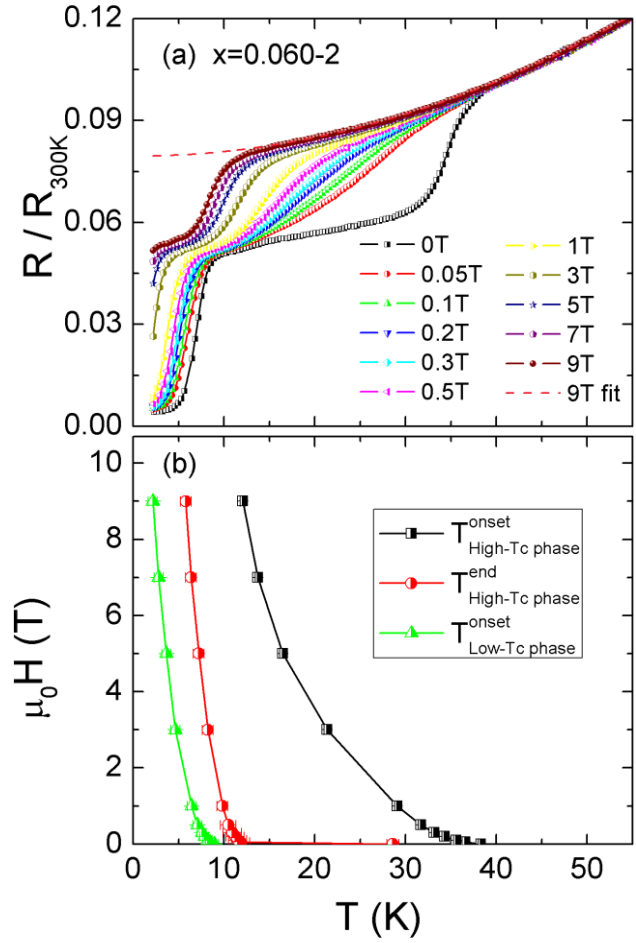

Figure. 4. (a) Temperature dependence of resistivity with applied fields up to 9T measured on one sample $x$=0.060-2. (b) Field-temperature phase diagram derived from the data in panel (a).

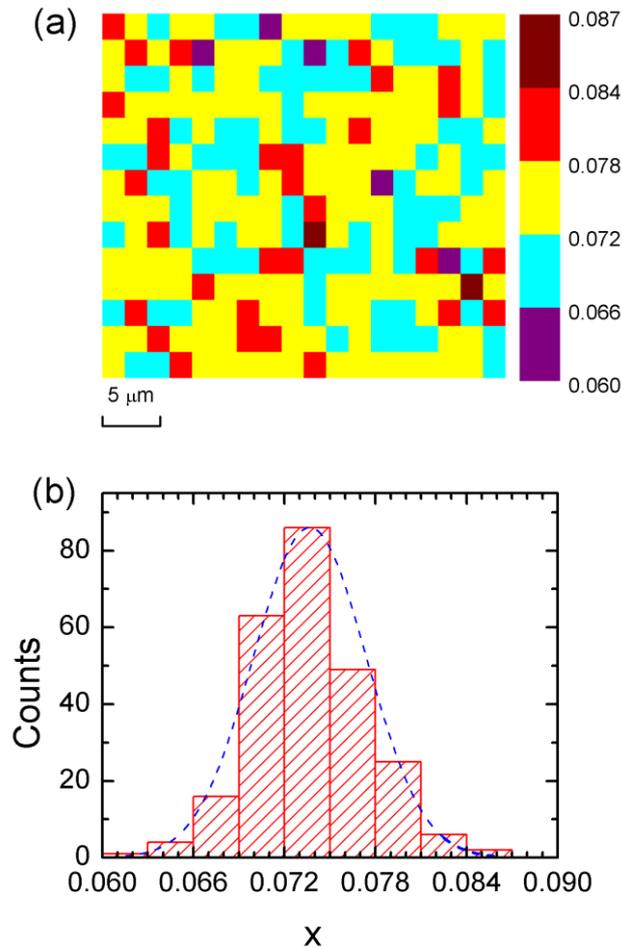

Fig. 5. The mapping image (a) of Nd concentration and chart of its distributions (b) throughout an area of 36×26 μm for the sample with $x = 0.074$, based on the EDS

quantitative results.

## 4 Conclusions

In the present work, we have investigated the phase diagram and field induced resistance broadening behavior of $Ca_{1-x}Nd_xFe_2As_2$ single crystals. It is found that $c$-axis lattice parameter decreases with the increase of Nd substitution. Coexistence of two SC phases is observed. The low-$T_c$ phase exists as $x \geq 0.031$, and the high-$T_c$ phase with a $T_c$ value of about 40 K emerges when $x \geq 0.060$. The structural (T-O)/magnetic transition is found in the low Nd-doping region and totally suppressed around $x = 0.060$. The new collapsed tetragonal structure coexists with the high-$T_c$ phase in the same doping range. Compared to the low-$T_c$ phase, the end point of the SC transition of high-$T_c$ phase shifts obviously to lower temperature even under a field of 0.05T, showing a weak-link behavior. Detailed EDS measurements indicate that the non-uniform SC properties cannot be attributed to the heterogeneous Nd distribution on the micro scale.

**Acknowledgements**

This work is supported by the Knowledge Innovation Project of Chinese Academy of Sciences (No. KJCX2-EW-W11), the Natural Science Foundation of China (No. 11204338), and the "Strategic Priority Program (B)" of the Chinese Academy of Sciences (No. XDB04040300).